\documentclass[doublespacing]{elsart_modified}

\usepackage{natbib}

\usepackage{graphicx}

\newcommand{\BibTeX}{ \textrm{B\kern-.05em\textsc{i\kern-.025em b}\kern-.08em
    T\kern-.1667em\lower.7ex\hbox{E}\kern-.125emX} }

\begin{document}

\begin{frontmatter}



\title{Composition of the L5 Mars Trojans: Neighbors, not Siblings}


\author[andy]{Andrew S. Rivkin\thanksref{visiting}},
\ead{andy.rivkin@jhuapl.edu} 
\author[dave]{David E. Trilling},
\author[mit]{Cristina A. Thomas\thanksref{visiting}}, 
\author[mit]{Francesca DeMeo},
\author[tim]{Timothy B. Spahr}, and
\author[mit]{Richard P. Binzel\thanksref{visiting}}  

\address[andy]{Johns Hopkins University Applied Physics Laboratory,
  11100 Johns Hopkins Rd. Laurel, MD 20723}
\address[dave]{Steward Observatory, The University of Arizona,
				Tucson, AZ 85721}
\address[mit]{Department of Earth, Atmospheric, and Planetary
  Sciences, M.I.T. Cambridge, MA 02139}
\address[tim]{Harvard-Smithsonian Center for Astrophysics, 60 Garden
  St., Cambridge MA 02138}
\thanks[visiting]{Visiting Astronomer at the Infrared Telescope Facility, which is operated by the University of Hawaii under Cooperative Agreement no. NCC 5-538 with the National Aeronautics and Space Administration, Science Mission Directorate, Planetary Astronomy Program.}



%
%
%
%


\end{frontmatter}


\begin{flushleft}
\vspace{1cm}
Number of pages: \pageref{lastpage} \\
Number of tables: \ref{lasttable}\\
Number of figures: \ref{lastfig}\\
\end{flushleft}


\newcommand{\sep}{; }


\pagebreak

\noindent
\textbf{Proposed Running Head:}\\
Characterization of Mars Trojans

\vspace{3cm}
\noindent
\textbf{Please send Editorial Correspondence to:} \\
Andrew S. Rivkin \\
Applied Physics Laboratory\\
MP3-E169\\
Laurel, MD 20723-6099, USA. \\
\\
Email: andy.rivkin@jhuapl.edu\\
Phone: (778) 443-8211 \\

\vfill

\pagebreak


\noindent
\textbf{ABSTRACT}


Mars is the only terrestrial planet known to have Trojan (co-orbiting)
asteroids, with a confirmed population of at least 4 objects.  The origin of
these objects is not known; while several have orbits that are stable
on solar-system timescales, work by Rivkin et al. (2003) showed they
have compositions that suggest separate origins from one another.  We
have obtained infrared (0.8-2.5 micron) spectroscopy of the two
largest L5 Mars Trojans, and confirm and
extend the results of Rivkin et al. (2003).  We suggest that the
differentiated angrite meteorites are good spectral analogs for
5261~Eureka, the largest Mars Trojan.  Meteorite analogs for
101429~1998~VF31 are more varied and include primitive achondrites and
mesosiderites.  

\vspace{\fill}
\noindent
\textit{Keywords:} Asteroids;\sep Asteroids, composition;\sep Trojan asteroids

\pagebreak

\section{Introduction}
Mars is the only terrestrial planet known to have co-orbiting
``Trojan'' asteroids.  These objects reside in stability zones
centered on the Lagrangian points 60$^{\rm o}$ leading and trailing
the planet. Jupiter and Neptune also have Trojan asteroid companions,
with the population associated with Jupiter estimated to number
roughly 160,000 with radii larger than 1 km in the L4 cloud.  By
contrast, Mars' known retinue is considerably more modest. Unlike
Jupiter Trojans, the definition of a Mars Trojan is not clear-cut,
with long-term integrations necessary to establish the stability of an
candidate's orbit.  At this writing, there are four
confirmed objects (listed below) and a handful of other suspects (including 2001~FR127)
identified by the Minor  Planet Center.  All but one of these objects
trail Mars.

The set of stable orbits near Mars has been studied by
\citet{tabachnik:marstrojans} and \citet{scholl:marstrojan}, who found
that 5261~Eureka, 101429~1998~VF31, 121514~1999~UJ7, and 2001~DH47 are in areas that are stable
on timescales longer than the age of the solar system, consistent with
an origin in their current orbits.  However,
\citet{rivkin:marstrojan} found the visible spectum of 1999~UJ7
to be significantly different from that of Eureka and 1998~VF31, and
that these objects were unlikely to have all formed at the same solar
distance.  The visible spectra and limited infrared spectrophotometry
suggested an Sa- or A-class membership for Eureka and 1998~VF31, and
an X-class membership for 1999~UJ7. Eureka is currently the largest known
Mars Trojan, with a diameter of roughly 1.3 km, while the diameter of 1998~VF31 is slightly
less than 800 m \citep{trilling:marstrojalb}.  A detailed discussion of the dynamics of Mars
Trojans can be found in \citet{scholl:marstrojan} and
\citet{connors:marscoorbitals},and for Trojans in general in
\citet{marzari:ast3}, but is beyond the scope of this paper.

\section{Observations}
Spectroscopic data were obtained using SpeX on the Infrared Telescope
Facility (IRTF) on Mauna Kea, Hawaii \citep{rayner:spex}.  The
instrument was in ``prism mode'', which provides continuous coverage
over the 0.8--2.5 $\mu$m spectral range with a resolution of 250.
Standard stars taken from the \citet{Landolt83} catalog and
extensively observed by \citet{bus:smass2data} were observed several times over the course of the nights
and are included in Table~\ref{obscirc}.  The
data reduction was performed using a combination of IRAF \citep{iraf}
and IDL routines to first extract the asteroid and standard spectra,
wavelength-calibrate the spectra, and finally to remove any residual
telluric contamination.  A more detailed
description of the data
reduction steps is provided in \citet{rivkin:hermes}.  The asteroid
5261~Eureka was observed on two nights, 1998~VF31 on
one.  The observing circumstances for each object are shown in
Table~\ref{obscirc}.

The spectra of silicates have been
shown to be temperature dependent in the wavelength region studied in this work
\citep{singer:temperature,hinrichs:temperature,moroz:temps},
potentially complicating comparisons to room-temperature laboratory
spectra. The 
objects presented in this work, however, have surface temperatures
near 250~K \citep{trilling:marstrojalb}.  Therefore, these
corrections are not important for Eureka and 1998~VF31 and in this
case can be safely neglected.




\begin{table}
\begin{center}
\begin{tabular}{llllll}
\hline 
\hline
Object & Date	& Observers	& V mag & Phase Angle	& Standard Stars \\
\hline
5261 Eureka	& 11 May 2005	& Binzel \& Thomas& 16.4 & 11.3$^{\rm o}$ &
L102,L105,L110 			\\
	& 19 May 2005	& Rivkin & 16.0 & 4.5$^{\rm o}$& L102, L105			\\
101429 1998 VF31	& 10 May 2005	& Binzel \& Thomas & 18.0 &
26.2$^{\rm o}$ &	L102, L110		\\
\hline
\end{tabular}
\caption[Observing Circumstances]	
	{
	\label{obscirc}	
	\label{lasttable}		
Observing circumstances for Mars Trojans. Codes for standard stars are
	as follows: L102 = L102-1081, L105=L105-56, L110=L110-361,
	where all are from the Landolt (1983) catalog.
	The V magnitudes for the asteroids were taken from the JPL
	Horizons ephemeris.
	}
\end{center}
\end{table}

\section{Results}

\subsection{5261~Eureka}
\begin{figure}[p!]
\begin{center}
\includegraphics[width=4in]{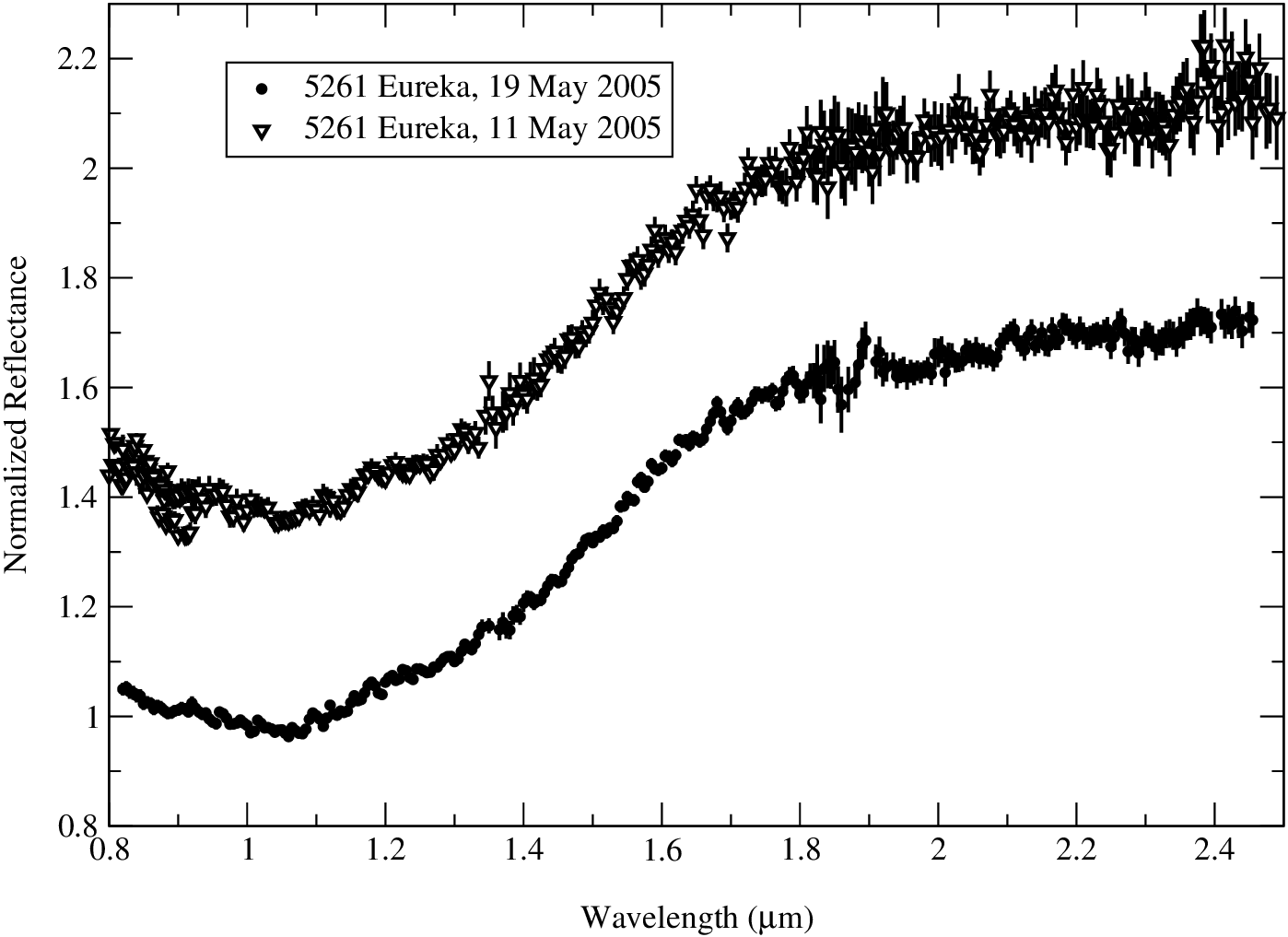}
\caption[Fit for Eureka]{
	\label{2eur}
	Two spectra of Eureka taken with Spex at the IRTF, from 11 and 19 May 2005.  The agreement
	between these spectra are excellent, both showing a broad,
	deep 1-$\mu$m absorption and a relatively straight, flat
	spectrum beyond $\sim$1.8 $\mu$m.  The spectra are normalized
	to 1 at 1 $\mu$m, and the 11 May spectrum was offset from the
	19 May spectrum for clarity.  The structure near 1.9 $\mu$m in
	the bottom spectrum is interpreted as noise rather than having
	mineralogical significance.
	}
\end{center}
\end{figure}

There is excellent agreement between the two Eureka spectra, as shown
in Figure~\ref{2eur}.  Because
the 19 May spectrum is of higher quality than the 11 May
spectrum (as expected since Eureka was brighter on that date), it
alone will be shown in later figures.  However, given the 
agreement between these spectra, all conclusions reached are consistent
with the first spectrum as well.

The spectrum of Eureka shows a broad, deep absorption band centered
near 1.08 $\mu$m, with no obvious corresponding 2-$\mu$m band.  This
is interpreted as evidence for olivine, with little if any
iron-bearing pyroxene present.  The visible-region spectra of Eureka
from \citet{rivkin:marstrojan} can be used to construct a full
0.4--2.5 $\mu$m spectrum, from which commonly-used spectral parameters
can be extracted \citep[see][for instance]{cloutis:calibrations}.


S-class and related asteroids are often interpreted using Band Area Ratio
(BAR)/Band~I plots,  popularized by
\citet{gaffey:stype}. In these plots, a subclassification within the S
class can be made using the position of the 1-$\mu$m band center and
the ratio of the areas of the 1- and 2-$\mu$m bands.  That
subclassification can give an idea of the relative proportions
of olivine and pyroxene on an object's surface as well as identiying potential
analogs among the known or a hypothesized
meteorite population.

As can be seen from Figure~\ref{2eur} and as mentioned above, there is no evidence for a resolvable
2-$\mu$m band on Eureka, leading to a BAR $\sim$ 0.  Its Band I
position is 1.08 $\pm$ 0.02.  Together, these lead to a classification in
the S(I) subclass.  \citet{gaffey:stype} suggests
that these objects are consistent with olivine-metal mixtures,
possibly analogous to pallasite meteorites. 

\begin{figure}[p!]
\begin{center}
\includegraphics[width=4in]{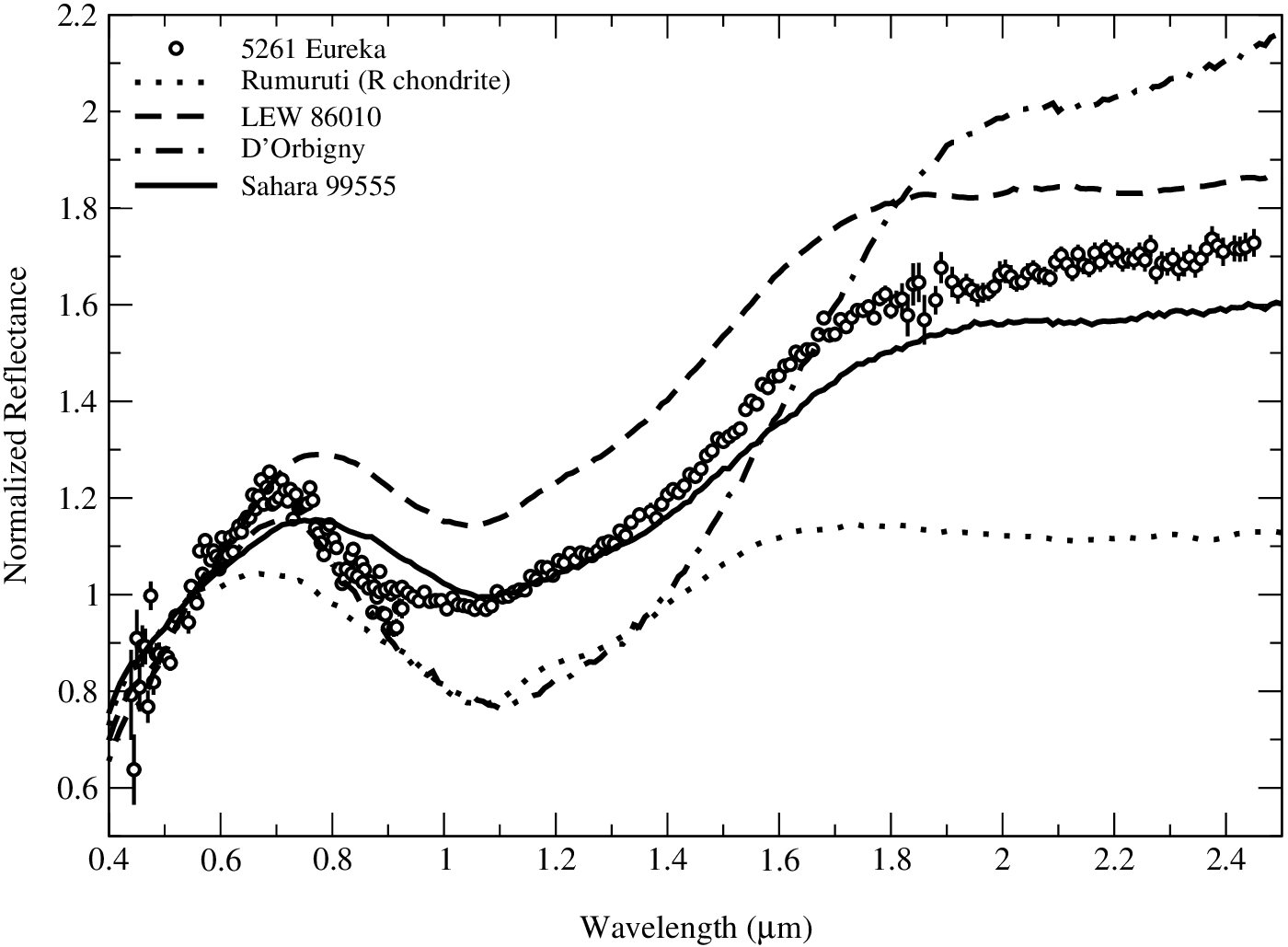}
\caption[Fit for Eureka]{
	\label{eureka_mets}
	Eureka compared to angrite meteorites, and also to the R
	chondrite Rumuruti.  The angrites bracket Eureka's spectrum
	well.  While Rumuruti is the only R chondrite for which a
	spectrum is available, it does not obviously match Eureka,
	failing to match its overall spectral slope.  However,
	additional spectra for additional R chondrites are necessary
	to fully reject them as analogs for Eureka.  
	}
\end{center}
\end{figure}

\citet{burbine:angritepaper} measured
reflectance spectra for 3 angrites and found them to have properties
similar to what we find for Eureka:  broad absorption bands near 1~$\mu$m and
weak or absent 2-$\mu$m bands, which would also be classified as S(I)
in the \citet{gaffey:stype} scheme.  Figure~\ref{eureka_mets} shows
Eureka's spectrum compared to the angrite spectra from
\citet{burbine:angritepaper}. The asteroid fits well among the others,
particularly LEW~86010 and Sahara~99555,
and is bracketed by the diversity shown in the meteorite
spectra.  

The angrite meteorites are igneous, generally basaltic rocks that have
unusual mineralogies including anorthosite, Ca-rich olivine, and
Ca-Al-Ti-rich pyroxenes. Their oxygen isotopes are similar to some
other groups, including the HEDs and
mesosiderites, but their peculiar mineralogies and compositions
suggest they are unrelated to those groups
\citep{weisberg:mess2}. Geochemical studies suggest that the angrites
are the igneous  products of carbonaceous chondrites, though other
origins have also been proposed \citep{kurat:dorbigny,varela:angrites}. 

\begin{figure}[p!]
\begin{center}
\includegraphics[width=4in]{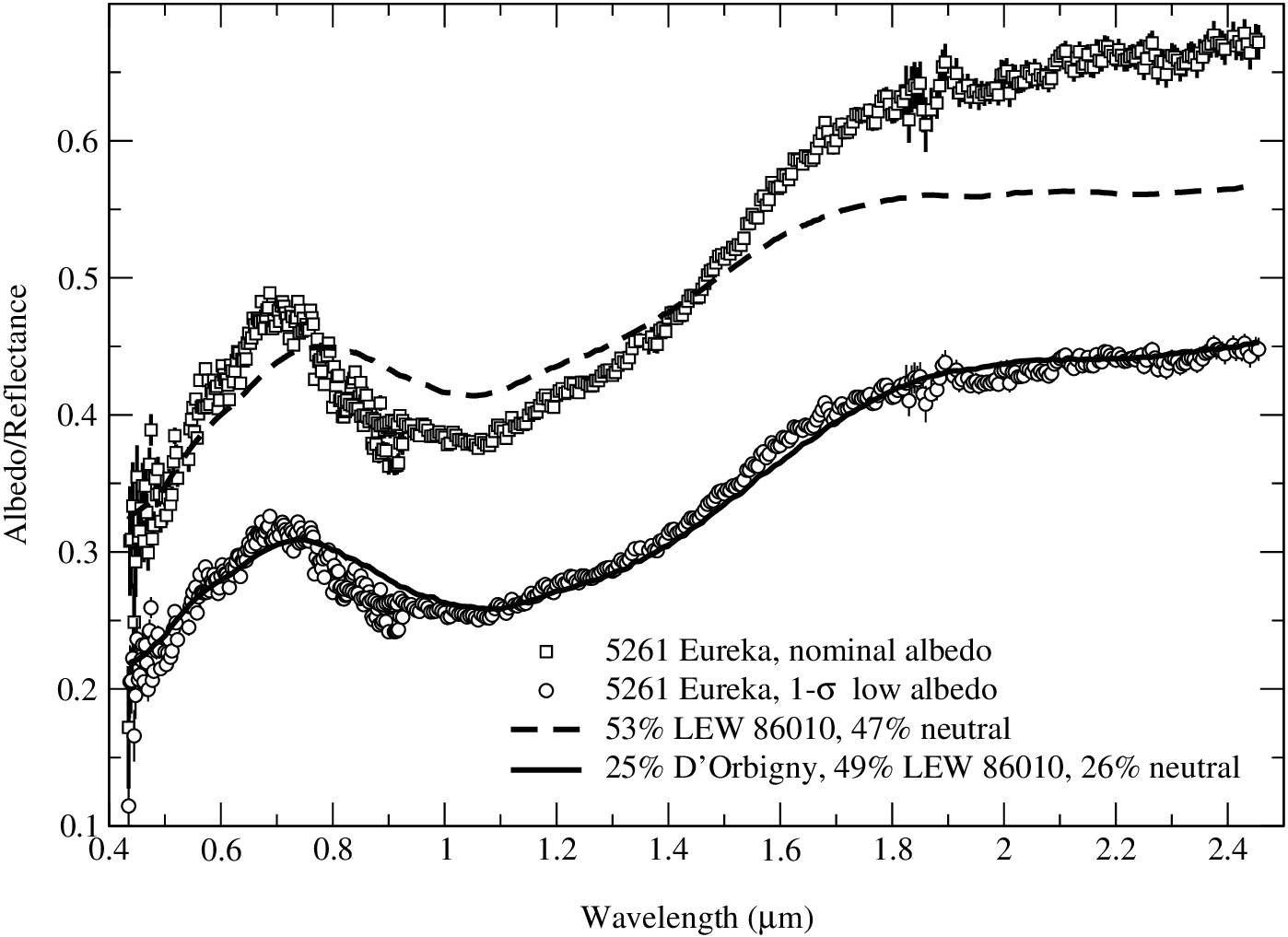}
\caption[Fit for Eureka]{
	\label{eureka_fits}
	Mixture modeling fits for Eureka.  These fits were calculated
	using only the angrites from the previous figure and flat
	(neutral) spectra as endmembers.  Two cases were modeled, one
	with an albedo of 0.39 for Eureka, the other 0.26.  These
	values match the mean and 1-$\sigma$ low values for Eureka's
	albedo, measured by Trilling et al. (in press).  The match is very good in
	particular for the lower-albedo case, and no
	other components are obviously necessary to provide an
	acceptable match.   

	}
\end{center}
\end{figure}

We modeled the reflectance spectrum of Eureka using angrites and
a neutral component to see if an acceptable fit could be achieved
using only those components.  The results are shown in
Figure~\ref{eureka_fits}.  We used a Hapke bi-directional scattering
model to mix end-member mineral spectra to simulate our whole-disk
observations of Eureka. Using software developed for analysis of NEAR
Shoemaker spectra of
asteroid 433 Eros \citep{clark:nearphot,clark:e-types}, we simulated intimate
mixtures of a set of 5 end-member spectra.  Our end-members were the
angrites D'Orbigny, Sahara~99555, and LEW~86010, and neutral phase
spectra with albedos varying from 0.1 to 0.6.
Spectrally neutral phases have been necessary in all spectral
modeling of airless rocky planetary surfaces to represent low-albedo
and/or non-crystalline components. End-member spectra were obtained
from T. Burbine for the angrites, or constructed for the neutral phases.
End-members were measured at grain sizes of less than 125 $\mu$m
(D'Orbigny and Sahara~99555) or less than 74 $\mu$m (LEW~86010).    Our
mixture model simulations are not meant to indicate unique determinations of
composition.  However, if the assumption is made that the end-member
choices are accurate, such models can be used as indicators of
relative proportions of detected components.  

The asteroid spectra were modeled including
albedo/reflectance information.  For the models, Eureka's albedo was
set to 0.39 or 0.26
representing nominal and 1-$\sigma$ low values of its measured
albedo. The determination of the radiometric albedos of Eureka and 1998~VF31 is
presented in a companion paper \citep{trilling:marstrojalb}. These albedo
values are consistent with the spectral class found for Eureka,
particularly NEOs of similar size observed by \citet{delbo:eta}.
As seen in Figure~\ref{eureka_fits}, Eureka's spectrum can be fit
quite well with only angritic and neutral components when using the
1-$\sigma$ low value for its albedo.  That fit
includes 55\% LEW~86010, with D'Orbigny contributing roughly half as
much to the spectrum and the neutral component (with albedo of
0.6) making up the remainder of the model fit components.  
Although not a unique fit, this excellent fit strongly supports an angritic
interpretation for Eureka. For the
nominal albedo case, the fraction of the high albedo neutral component
must increase, and does so at the expense of D'Orbigny, the
lower-reflectance angrite.  The fit is poorer than the low-albedo
case.  The 1-$\sigma$ high albedo for Eureka does not result in an
acceptable fit, and is not shown.  
 
Another possible match is also shown in Figure~\ref{eureka_mets}.
Eureka qualitatively matches a spectrum of Rumuruti powder
\citep{burbine:eros,sunshine:oxygen}, a member of the rare R
chondrite class, and the only known fall of that type.  R chondrites
are relatively oxidized, anhydrous, metal poor, primitive meteorites
with high olivine fayalite contents \citep{mcsween:mess2}.  They are thought to
have formed at greater heliocentric distances than the ordinary
chondrites \citep{kallemyn:rchonds}.  In this interpretation, Eureka
would be undifferentiated.  However, we prefer the angrite
interpretation for Eureka for two reasons:  first, the match to Eureka in the
blue and ultraviolet region is better for the angrite spectra than for
Rumuruti.  Second, spectra of Rumuruti vary considerably, with those
of saw-cut surfaces showing blue spectral slopes for two of the three
lithologies measured and a too-low albedo for the third lithology
\citep{berlin:rumuruti}.  We note that spectra from these saw-cut
surfaces may not
be a good representation of the spectra one would receive from an
asteroidal surface.  Additional spectra of Rumuruti and other R chondrites
may strengthen or weaken an R-chondrite interpretation of Eureka
with respect to possible meteorite parent bodies.
However, at this time we cannot formally rule out a connection
between Eureka and the R chondrites.

\subsection{101429~1998 VF31:}
\begin{figure}[p!]
\begin{center}
\includegraphics[width=4in]{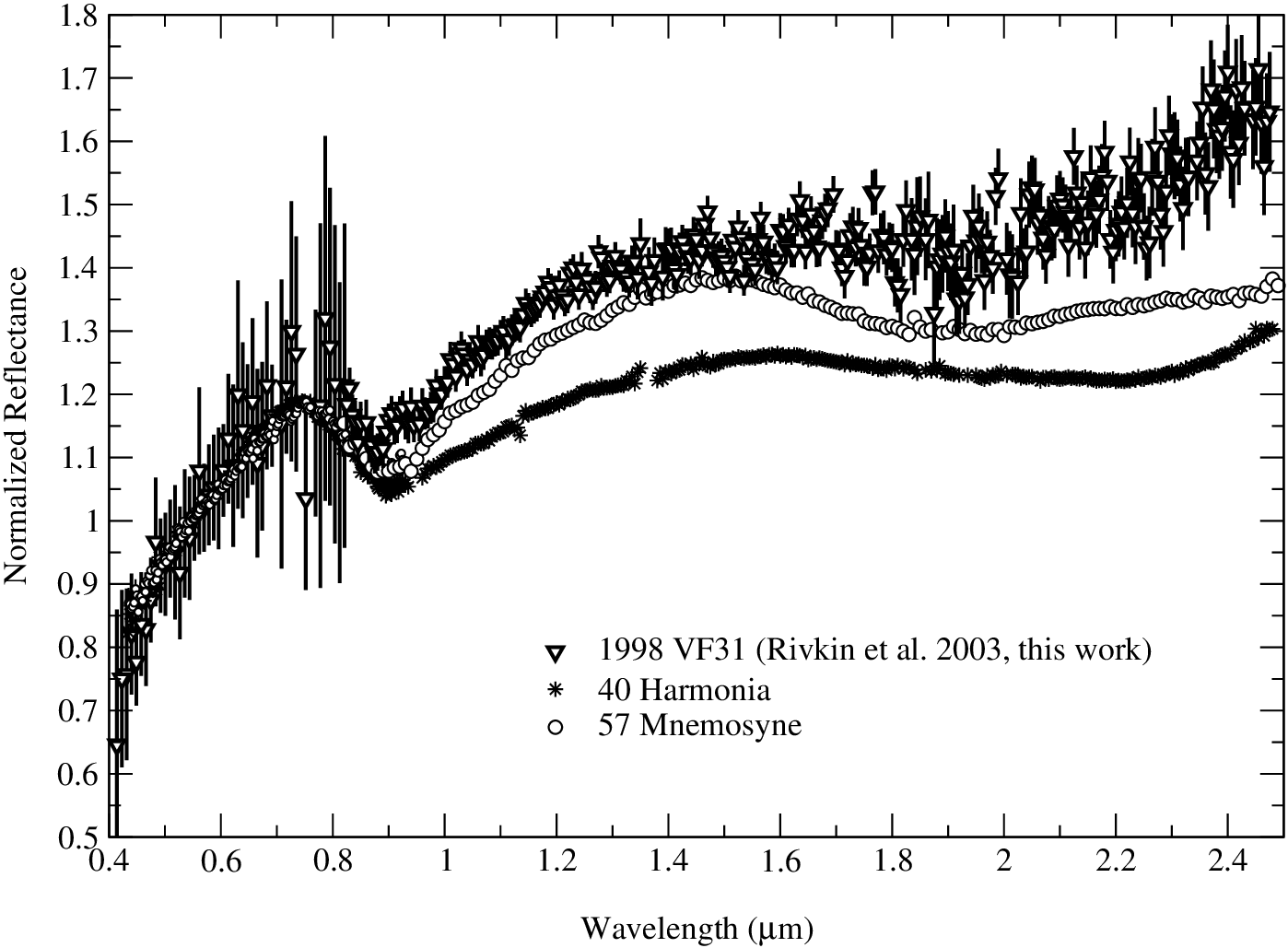}
\caption[1998 VF31]{
	\label{vf31a}
	Spectrum of 1998~VF31 compared to two S-class asteroids.  All
	are classified as S(VII) objects in the Gaffey et al. (1993)
	subclassification scheme.  The spectral slope for 1998~VF31 is
	higher than for either main-belt asteroid, but the overall
	spectral characteristics are qualitative matches.  The rise in
	reflectance for 1998~VF31 beyond $\sim$2.2 $\mu$m is not due
	to thermal emission, and can be seen in 40~Harmonia, as well.   

	}
\end{center}
\end{figure}

The asteroid 1998~VF31 has a visible and near-IR spectrum typical of
S-class asteroids (Figure~\ref{vf31a}). Although the data are rather
noisier than the Eureka spectrum, we can still determine the band
parameters, with a band center of 0.9 $\pm$ 0.03 and a BAR of 2.0
$\pm$ 0.3.  While not as precise as might be desired, this is still
sufficient to firmly place 1998~VF31 into the S(VII) subclass.
Figure~\ref{vf31a} includes two bright S(VII)
asteroids (40~Harmonia and 57~Mnemosyne) for which high-quality SMASS
and Spex data are available, showing a
similarity that supports such a classification for 1998~VF31.  

In \citet{rivkin:marstrojan}, 1998~VF31 was interpreted as an Sr, Sa,
or A-class object based on its similarity to the spectrum of Eureka
over the limited visible wavelengths that it was observed.  With the
benefit of much-expanded wavelength coverage and a near-IR spectrum of
higher quality than the visible spectrum previously available, it is
clear that 1998~VF31 is quite different from Eureka beyond 0.8
$\mu$m.  Its spectrum shows both 1- and 2-$\mu$m bands, typical of
S-class asteroids.  

An upturn in flux beyond 2.1 $\mu$m for 1998~VF31 is reminiscent of
thermal emission seen in some near-Earth objects.  While the surface
temperatures of Mars Trojans are warmer than their main-belt cousins,
the albedo required for detectable thermal emission near 2.4-2.5
$\mu$m at 1998~VF31's solar distance is unrealistically low ($\sim$ 0.04 or less) given what
is usually seen for S-class asteroids and would conflict with the value
of $\sim$0.32 measured by \citet{trilling:marstrojalb}.  Therefore, we
interpret the
upturn as a part of a general spectral slope on the object.
Interestingly, the much larger main-belt asteroid 40~Harmonia also has
an upturn at longer wavelengths, as seen on Figure~\ref{vf31a}. 

The S(VII) asteroids are interpreted by \citet{gaffey:stype} as
possibly analogous to any of a variety of primitive achondrites, or
possibly mesosiderites, though they found that interpretation
unlikely.  Heating of a chondritic precursor in the presence of a
reducing agent was also proposed as a possibility.

\begin{figure}[p!]
\begin{center}
\includegraphics[width=4in]{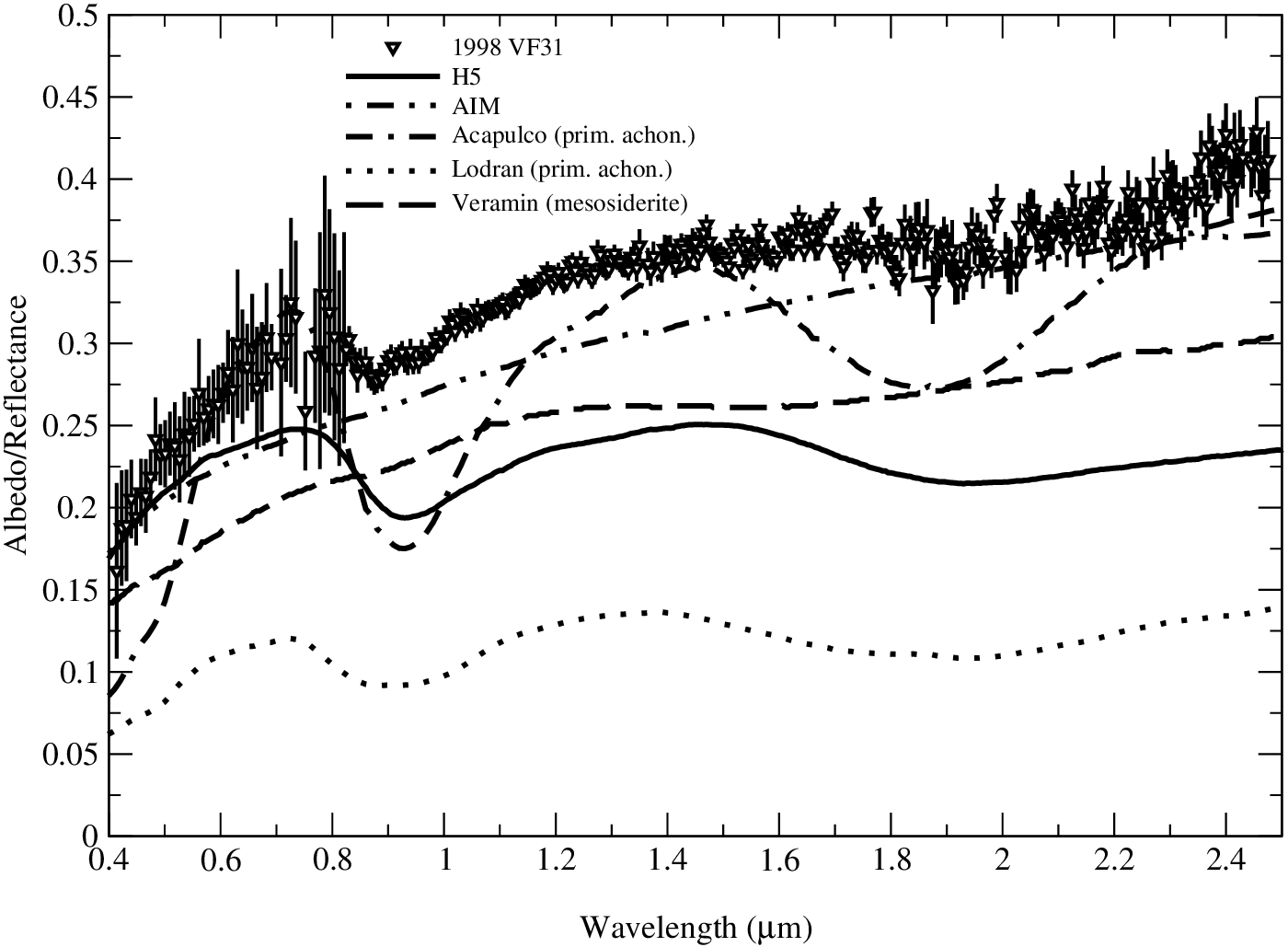}
\caption[1998 VF31 endmembers]{
	\label{vf31end}
	Endmembers used in mixing models for 1998~VF31.  The
	mesosiderite Veramin, the average
	spectra for H5 meteorites and iron meteorites (AIM in the
	figure) were used from Gaffey (1976).  The primitive
	achondrite spectra were obtained from RELAB. The spectrum of
	1998~VF31 is also included, scaled to the nominal measured
	albedo from Trilling et al. (in press).  Primitive achondrites and
	mesosiderites are possible analogs for S(VII) asteroids
	according to Gaffey et al. (1993).  Ordinary chondrites were
	also included in the mixing models, though only H chondrites
	had a contribution to the fit.  All of these meteorite
	spectra are of lower albedo than the asteroid, as with Eureka.

	}
\end{center}
\end{figure}

\begin{figure}[p!]
\begin{center}
\includegraphics[width=4in]{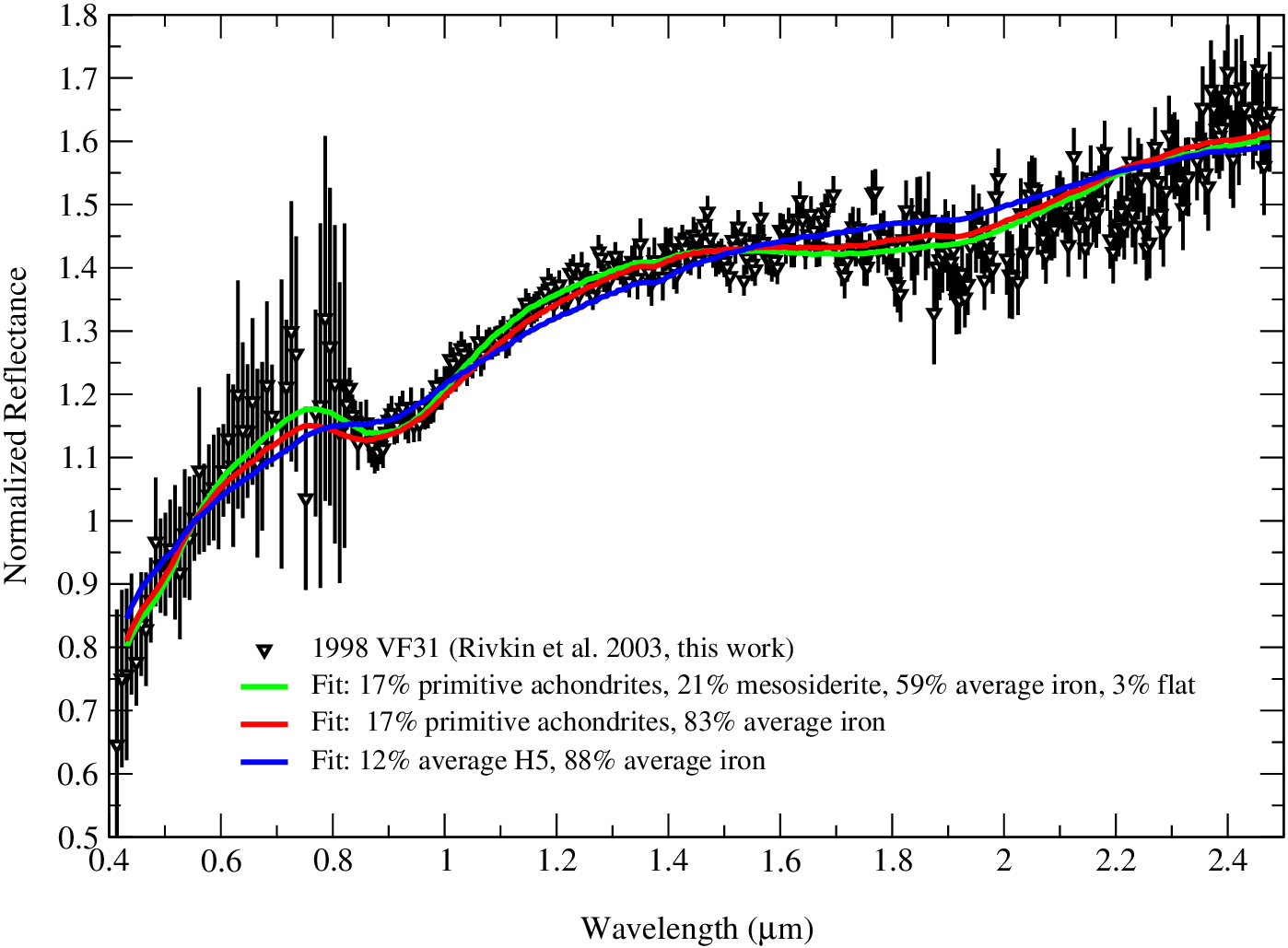}
\caption[1998 VF31 mix]{
	\label{vf31mod}
	\label{lastfig}			
	Mixture Models for 1998~VF31.  Although the spectrum for
	1998~VF31 is noisier than for Eureka, the mixture model is
	still instructive. The model assemblage is dominated by iron,
	with roughly 17\% primitive achondrite.  Models with ordinary
	chondrite endmembers are even more dominated by iron, and do
	not produce as good a fit.  All of these model fits have
	albedos of roughly 0.2, consistent with the 1-$\sigma$ low albedo of
	1998~VF31 from Trilling et al.(in press).
	}
\end{center}
\end{figure}

As with Eureka, mixing models were performed for 1998~VF31, using the primitive
achondrites Acapulco and Lodran, the mesosiderite Veramin, the average
of the iron meteorites (AIM) in \citet{gaffey:metspec}, and a neutral
component.  The meteorite endmembers are shown in
Figure~\ref{vf31end}.  The larger uncertainties in 1998~VF31's
spectrum, as well
as a smaller selection of suitable endmember spectra, leave this fit
somewhat less certain than the fit for Eureka.  As shown in
Figure~\ref{vf31mod}, the model spectra are dominated by metal, with
roughly 80\% of the contribution coming from a combination of the AIM and
Veramin spectra, with the remaining 20\% contribution from a combination of
the primitive achondrites.  A fit using only H5 chondrite ($\sim$12\%, also
shown in Figure~\ref{vf31mod}) and AIM
($\sim$88\%) spectra is also consistent, though is of much lower
quality.  All of these fits give an albedo of roughly 0.2, again lower
than the nominal radiometric albedo of 0.32 $^{\rm +0.18}_{\rm -0.11}$  in
\citet{trilling:marstrojalb}
but consistent within observational uncertainties.  It is conceivable that
the large metal fraction indicated by these fits is a sign that the
surface of 1998~VF31 has undergone regolith maturation \citep[aka
  space weathering:][and others]{clark:ast3}, but we might expect
little regolith on an asteroid of this size. A more thorough
consideration of space weathering and 1998~VF31 is beyond the scope of
this work.  However, we conclude that 1998~VF31's spectrum is most
consistent with a mixture of metal and primitive achondrites, either
with or without mesosiderite contribution.  


\section{The origin of the Mars Trojans}


The visible spectra of three Mars Trojans were interpreted by
\citet{rivkin:marstrojan} as showing that they all could not have
formed at their current distances, though the two objects at L5
(Eureka and 1998~VF31) could be related.  With expanded wavelength
coverage and the most likely mineralogical interpretations presented
above, it appears unlikely that the L5 objects are related to one another.  An
angritic composition for Eureka suggests an oxidized, carbonaceous
chondritic precursor.  If Eureka is more like Rumuruti and the R
chondrites, it would again be expected to be more oxidized than the
ordinary chondrites.  1998~VF31, if a primitive achondrite, would be
expected to have originated on reduced objects relative to ordinary
chondrites. Furthermore,all of the suggested analogs for S(VII)
objects have significantly different oxygen isotopes than angrites or
the R chondrites, again indicating a different parent body (or, of
course, that we don't have samples of the relevant objects).  The most
straightforward explanation is that Eureka and 1998~VF31 are not
related, and presumably at least one (if not both) were captured into
their current orbits at some point in solar system history, perhaps
very early. We also note for completeness that Phobos and
Deimos have spectra very different from the L5 objects, suggestive of
outer-belt asteroids or mature lunar soils
\citep{rivkin:phobos,gendrin:phobos}, as discussed further in
\citet{rivkin:marstrojan}.   

New dynamical models have been proposed in recent years that suggest a
large flux of objects were scattered from the asteroid belt and Kuiper
belt roughly 4 billion years ago
\citep{gomes:nicemodel}.  It has been proposed that during this
period, the Trojan asteroids of Jupiter were captured into their
current orbits \citep{morbidelli:trojans}.  Morbidelli
\citep[private communication, also reported in][]{scholl:marstrojan}
suggested that the capture of Mars Trojans could be aided by a chaotic
wandering of Mars' semi-major axis due to impacts between proto-Mars
and lunar-size planetary embryos.

The igneous events that formed the angrites occurred very early in
solar system history, 
allowing plenty of time for collisions to break up the angrite parent
body and then allow at least one piece to be captured as a Mars
trojan.  Argon dating of R chondrites by \citet{dixon:rchond} suggests
early impact events on that parent body, again creating the
opportunity for ejecta to be created and captured into Mars Trojan orbits.  

Given the much larger set of possible compositions for 1998~VF31, its
story is less easy to pin down. Mesosiderites are thought to have had
a major degassing event roughly 3.9 billion years ago, interpreted as
collisional disruption \citep{rubin:mesosiderite} and consistent
with capture by Mars at that time.  This is also the same era as the
late heavy bombardment.  Primitive achondrites are thought
to have been heated early in solar system history, as well.  

Taking the spectroscopic, meteoritic, and dynamical evidence all
together, the simplest explanation for the origins of Eureka
and 1998~VF31 is that they formed separately in other parts of the inner solar
system as part of larger bodies, those bodies were disrupted during
the earliest times in the solar system and pieces found themselves
trapped in the 1:1 resonance with Mars by roughly 3.9 billion years
ago. Thus, they are most likely to be long-term residents but not
natives.  Additional work will be necessary to determine the total
population of the neighborhood and whether all of the inhabitants are
immigrants.

\ack
The data taken here were obtained through the SMASS and MIT-IRTF-UH
surveys as targets of opportunity.  This work was supported by
NASA Planetary Geology and Geophysics grant NNG06GA23G. As usual,
Dave, Bill, Paul, and Eric at the IRTF were
indispensible to actually getting the data in the first place. Data
from RELAB were critical for our analysis, and we thank Tom Burbine
for sharing his meteorite spectra.  Helpful reviews by Tom Burbine and Sonia
Fornasier improved and strengthened this manuscript.  This
work has made use of NASA's Astrophysics Data System.  Thanks to
Gwen Bart, Ross Beyer, Dave O'Brien, and Paul Withers
for creating the \LaTeX template used for this manuscript. And continuing thanks to the indigenous people of Hawai'i for allowing astronomers to use their sacred mountain.  

\label{lastpage}


\bibliography{bibliography.bib,../refs,../asti,../uapress,../ast2,../refs2,../3mic,../ast3,../mess2}
\bibliographystyle{elsart-harv}


\clearpage	



\clearpage


\end{document}